\documentclass[a4paper,fleqn]{cas-sc}

\usepackage[authoryear]{natbib}

\usepackage[shortlabels]{enumitem}
\usepackage{amsmath}
\usepackage{comment}
\usepackage{tikz} 
\usetikzlibrary{matrix}
\usepackage[]{graphicx}
\graphicspath{{Figures/}}
\usepackage{float}
\usepackage{siunitx}

\usepackage[english]{babel}
\usepackage[T1]{fontenc}
\usepackage{etoolbox}
\makeatletter
\patchcmd{\ps@pprintTitle}{\footnotesize\itshape
      Preprint submitted to \ifx\@journal\@empty Elsevier
      \else\@journal\fi\hfill\today}{\scriptsize{Preprint submitted to Computers & Geosciences \hfill \today}}{}{}

\def\ps@pprintTitle{%
  \let\@oddhead\@empty
  \let\@evenhead\@empty
  \let\@oddfoot\@empty
  \let\@evenfoot\@oddfoot
}
\makeatother
\usepackage{comment}

\usepackage{subfig}
\usepackage{hyperref}
\usepackage{blindtext}
\usepackage{tcolorbox}
\usepackage{amsmath}

\usepackage{amsmath,amssymb,amsfonts}
\usepackage{algorithmic}
\usepackage{lipsum}
\usepackage{balance}
\usepackage{textcomp}
\usepackage{xcolor}
\usepackage{multirow}
\usepackage{color}
\usepackage{lscape}
\usepackage{graphicx}
\usepackage{enumitem}
\usepackage{booktabs,siunitx,rotating}

\usepackage[switch]{lineno}

\usepackage{array}

\def\tsc#1{\csdef{#1}{\textsc{\lowercase{#1}}\xspace}}
\tsc{WGM}
\tsc{QE}
\tsc{EP}
\tsc{PMS}
\tsc{BEC}
\tsc{DE}

\usepackage{lineno}

\usepackage{nomencl}
\makenomenclature

\begin{document}
\let\WriteBookmarks\relax
\def\floatpagepagefraction{1}
\def\textpagefraction{.001}
\shorttitle{Link Climate}
\shortauthors{Wu et~al.}

\title [mode = title]{Link Climate: An Interoperable Knowledge Graph Platform for Climate Data}



\author[1,2]{Jiantao Wu}[]
\ead{jiantao.wu@ucdconnect.ie}

\author[1,3]{Fabrizio Orlandi}[]
\ead{fabrizio.orlandi@adaptcentre.ie}

\author[1,3]{Declan O'Sullivan}[]
\ead{declan.osullivan@adaptcentre.ie}

\author[1,2]{Soumyabrata~Dev}[type=editor,auid=000,bioid=1,orcid=0000-0002-0153-1095]
\ead{soumyabrata.dev@ucd.ie}
\ead[URL]{https://soumyabrata.dev/}
\cormark[1]

\address[1]{The ADAPT SFI Research Centre, Ireland}
\address[2]{School of Computer Science, University College Dublin, Ireland}
\address[3]{School of Computer Science and Statistics, Trinity College Dublin, Ireland}

\cortext[cor1]{Corresponding author}


\begin{abstract}
Climate science has become more ambitious in recent years as global awareness about the environment has grown. To better understand climate, historical climate(\textit{e.g.} archived meteorological variables such as temperature, wind, water, etc.) and climate-related data (\textit{e.g.} geographical features and human activities) are widely used by today's climate research to derive models for an explainable climate change and its effects. However, such data sources are often dispersed across a multitude of disconnected data silos on the Web. Moreover, there is a lack of advanced climate data platforms to enable multi-source heterogeneous climate data analysis, therefore, researchers must face a stern challenge in collecting and analyzing multi-source data. In this paper, we address this problem by proposing a climate knowledge graph for the integration of multiple climate data and other data sources into one service, leveraging Web technologies (\textit{e.g.} HTTP) for multi-source climate data analysis. The proposed knowledge graph is primarily composed of data from the National Oceanic and Atmospheric Administration's daily climate summaries, OpenStreetMap, and Wikidata, and it supports joint data queries on these widely used databases. This paper shows, with a use case in Ireland and the United Kingdom, how climate researchers could benefit from this platform as it allows them to easily integrate datasets from different domains and geographical locations.
\end{abstract}




\maketitle

\mbox{}
\nomenclature{\(CA\)}{Climate Analysis Ontology}
\nomenclature{\(KG\)}{Knowledge Graph}
\nomenclature{\(NOAA\)}{National Oceanic and Atmospheric Administration}
\nomenclature{\(CDO\)}{NOAA Climate Data Online}
\nomenclature{\(RDF\)}{Resource Description Framework}
\nomenclature{\(OSM\)}{OpenStreetMap}
\nomenclature{\(SSN\)}{Semantic Sensor Networks Ontology}
\nomenclature{\(SOSA\)}{Sensor, Observation, Sample, and Actuator Ontology}
\nomenclature{\(AEMET\)}{the Spanish Agency of Meteorology}
\nomenclature{\(OWL\)}{the Web Ontology Language}

\printnomenclature[1in]

\section{Introduction}
Climate change is a pressing and urgent topic in today's climatic and meteorological research. Scientists are working nonstop to find ways to avoid the harmful consequences of climate change in the future. One of the most important challenges being investigated by researchers is currently how to model big historical climate data and develop algorithms based on the data to predict the effects of climate change~\citep{Hu2018-ea, Yu2020-ix, Poppick2020-yd, fathima2021role, fathima2019chaotic}. 
However, climate can be understood to be a complex ensemble of various meteorological variables such as cloud, precipitation, temperature, wind, water, and so on, all of which are critical to furthering the scientific knowledge of climate~\citep{Manandhar2019-at, manandhar2018systematic, wu2022boosting, pathan2021analyzing,dev2017cloud}. 
The resulting analytical model is inextricably linked to the input data, which consists of observations of the meteorological variables mentioned above. This usually results in time-consuming attempts to identify a sufficient range of suitable input datasets~\citep{wu2021interoperable,wu2021uplifting, wu2022kgml}. Additionally, although modern models, particularly those employed in automation systems, are capable of digesting large amounts of data and reaching a certain degree of performance, it is difficult to ascertain the reason behind the automated decisions~\citep{Lacayo2021-dd, wu2021detecting, wu2021automated}. In these intelligent platforms, the way models organize data is much more machine-readable than human-readable.

On the other hand, climate as a system influences - and is also influenced by - additional elements, \textit{e.g.} regional characteristics~\citep{martel2018role}, 
socioeconomic variables such as gross domestic product, transportation routes, agricultural, and industrial production~\citep{bojinski2014concept}. There is still a large room for improvement by adding new datasets (\textit{e.g.} satellite and remote sensing data) to the existing models~\citep{Liu2021-sg}. 
The investment on finding new and diverse (multi-dimensional) data sources is becoming larger~\citep{Beretta2021-ew, wu2021organizing,wu2021dec}. As yet many climatic studies are facing a challenge in examining the relationships between the climate domain and others~\citep{bonan2018climate,wu2022semantic,wu2022pca,wu2022awc}. 
Some of the difficulties may stem from the lack of frameworks or platforms that facilitate cross-domain analysis. Perera et al.~\citep{perera2018quantifying}, for example, is adamant about the need for a powerful computational platform that can assess a mix of urban climate, building modeling, and energy system optimization all at once to improve energy system performance. When distinct scientific aims are identified, meteorological variables are supposed to be allocated varying relative importance according to standard criteria, as highlighted in a recent survey~\citep{burrows2018characterizing}. According to Burrows et al., new research methodologies for studying climate effects on other disciplines are in urgent need of development.

In view of the growing need for meteorological data sources and computing platforms, this study examines how contemporary Knowledge Graph (KG) approaches can be used as an answer to these difficulties in the climate area. Specifically, we present a standards-based methodology for creating a KG and dynamically updating it using historical climate data from climate data providers (daily observations of meteorological variables and stations).  
In addition, we show how to integrate geographical information about weather stations from OpenStreetMap\footnote{\url{https://www.openstreetmap.org/}} and enrich this data by linking identical locations to the corresponding ones on Wikidata (a Wikipedia-based open KG). As a result, our methodology based on KGs facilitates the simultaneous study of numerous meteorological variables as well as cross-domain analysis of heterogeneous climate datasets. The approach is based on W3C Semantic Web standards\footnote{\url{https://www.w3.org/standards/semanticweb/}}~\citep{Berners-Lee2001-rx} and Linked Data principles~\citep{bizer2011linked}. 
As a demonstrating example, along with Wikidata\footnote{\url{https://www.wikidata.org/}} and OpenStreetMap, we use the NOAA (National Oceanic and Atmospheric Administration) climate data \footnote{\url{https://www.ncdc.noaa.gov/cdo-web/}}.  

The rest of the paper\footnote{In the spirit of reproducible research, the source code is available at \url{https://github.com/futaoo/LinkClimate}.} is organized as follows: Section~\ref{sec:relatedwork} introduces the related work and indicates where our contributions stand in comparison to said work. Then we provide an overview of the whole workflow in Section~\ref{sec:graphical overview} and describe in Section~\ref{sec: design} how each component involved contributes to the KG platform's functionality. Finally, in Section~\ref{sec:web interface} we examine the usability study of our platform to demonstrate how the platform is evaluated by potential users. We suggest that the proposed platform should be considered as a starting point of climate-relevant KG research, and that other KG applications in the climate  area should be investigated further.

The novel contributions of this paper are as follows:
\begin{itemize}
    \item An open online KG populated with NOAA climate data as a means of providing context to data, thus increasing the platform's explainability, which is often lacking in many automation systems.
    \item Integration of heterogeneous data sources \textit{e.g.} climate, with geographic (OpenStreetMap) and encyclopedic (Wikidata) source through use of Linked Data Principles.
    \item Regular, automated synchronization of heterogeneous data into the KG.
    \item A Web interface to assist climate researchers in exploring and using the platform.
\end{itemize} 

\section{Related Work}
\label{sec:relatedwork}
The proposed work spans across the following distinct but related domains: ontology modeling (knowledge engineering), geographic data integration, Linked Data, KGs and RDF (Resource Description Framework) data integration and publication. 
In this section, we describe how prior work in each domain makes up our contributions.

One of the fundamentals to create a KG is to take advantage of ontologies to semantically structure the graph data. Especially for sensor-related data (\textit{e.g.} climate observations), SSN (the Semantic Sensor Networks)~\citep{haller2019modular} and SOSA (Sensor, Observation, Sample, and Actuator)~\citep{janowicz2019sosa} ontologies are widely employed by scholars as the basic semantics to create domain-specific KGs and for defining and representing data. For example, in our previous work we have proposed the Climate Analysis (CA) ontology~\citep{wu2021ontology}, an extension of the SOSA ontology by adding adaptable and concrete meteorological and geographic elements associated terms designed to semantically represent tabular NOAA climate data. 
In this work, we enhance the CA ontology with additional vocabularies (see Section~\ref{section: ontTrans}) to further support the flexible API-based NOAA climate data which will be used to create a climate  KG able update with dynamic NOAA climate data. In addition, we also align our ontology with additional classes and properties of the AEMET (the Spanish Agency of Meteorology) ontology, which is also defined for the climate domain by Vilches-Bl{\'a}zquez~\citep{vilches2014integrating}.  
The aim of alignment is to improve the interoperability of heterogeneous resources.

Integration of geographic data into climate domain has been popularly investigated by many researchers. The common tendency is to use GeoNames~\citep{wick2015geonames}, which is a derived knowledge base of Google Maps, as the source for geographic data integration~\citep{escobar2020ontology}. 
However, there are some other competitive volunteered geographic information providers, but few to be investigated for geographic data integration in climate  KGs. For example, OpenStreetMap (OSM) is recently evolved as a key backbone for various OSM applications such as spatial data mining, geographic information services\footnote{\url{https://histosm.org/}}, and GPS tracking\footnote{\url{https://gitlab.com/eneiluj/phonetrack-oc}}. The variety of application innovations create large potential in widening the climate research horizon. Moreover, the geographic entities in OSM (so-called ``OSM nodes'') have been linked to Wikidata and DBpedia to bring more contextual, semantic information~\citep{tempelmeier2021linking}. This motivates our work in using OSM to include geographic information, as well as contextual information for spatial nodes, in our climate KG (\textit{e.g.} weather stations) instead of GeoNames.

Linked Data, Semantic Web and RDF data publication have been proposed in the past to make KGs available for users across the Web. We use Apache Jena\footnote{\url{https://jena.apache.org/documentation/fuseki2/}} Fuseki~\citep{seaborne2011fuseki} \textsuperscript{TM} to configure a triple-store (a RDF graph database) and open a SPARQL endpoint on the Web for users to query our climate  KG.  
In addition, we follow the Linked Data principles to construct the KG and every node is associated with a unique URI able to be dereferenced with any standard web browser. We adopted the tool LodView\footnote{\url{https://lodview.it/}} to dereference these URIs. The advantage of using LodView is that it is a graph-like RDF browser and the dereferenciation complies with W3C standards.

\section{Workflow}
\subsection{Graphical Overview}
\label{sec:graphical overview}
\begin{figure}[ht]
\centering
\includegraphics[width=0.8\textwidth]{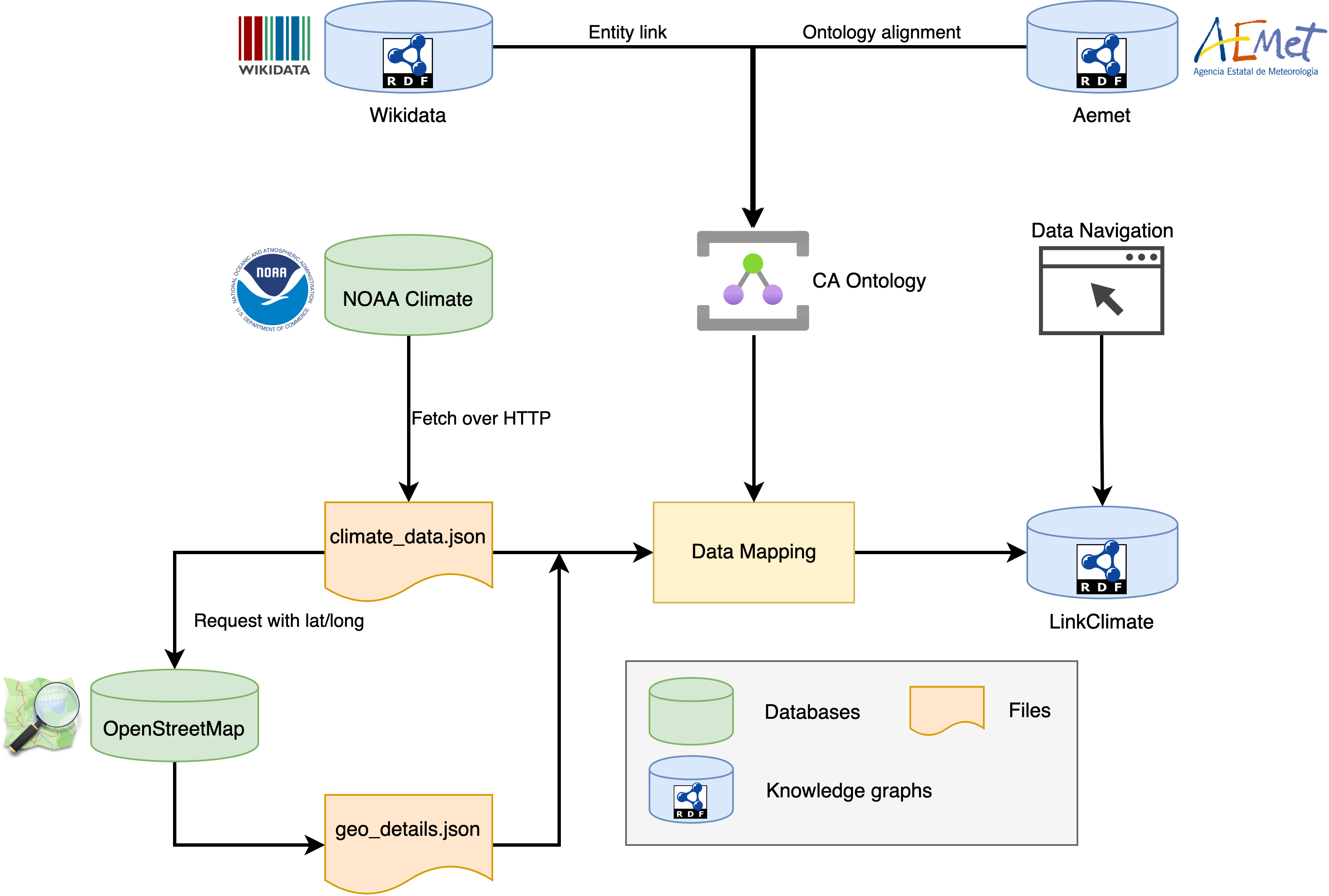}
\caption{A graphical illustration of the whole workflow}
\label{fig:workflow}
\end{figure}

 Fig~\ref{fig:workflow} depicts an automatic procedure to integrate and adapt climate  and ``beyond-climate'' data within KG's norms.  
 First, the procedure starts by requesting climate data from the NOAA service APIs (described in Section~\ref{sec:selection of NOAA data}) as an input for the KG's automated creation. It should be noted that a KG is meant to be assembled following specific KG data schema \textit{aka.} ``Ontology''~\citep{hogan2020knowledge}. The CA ontology (described in Section~\ref{section: ontTrans}) is used to structure and introduce consistent semantics to the raw data retrieved from NOAA. Second, OpenStreetMap-based geographic information, such as counties and cities of climate stations, is integrated to enrich the KG serving as the reference for including the corresponding identical entities in Wikidata (specified in Section~\ref{sec: linkage to wiki}). 
 Finally, we use a Linked Data platform\footnote{\url{http://jresearch.ucd.ie/linkclimate/index.html}} to publish the KG and dereference all URIs of entities and concepts presented in the CA ontology (specified in Section~\ref{sec:uri deref}) for data navigation over HTTP. Further details about each step are presented independently in the following workflow design section.

 \subsection{Design of the workflow components}
 \label{sec: design}
 
 \subsubsection{Selection of NOAA data services}
 \label{sec:selection of NOAA data}
 
 The application is designed to get climatic data from the data vendor as input at the start of the workflow. According to the NOAA website, there are three main options for fetching the  data and for implementing this initial step: Search Tool\footnote{\url{https://www.ncdc.noaa.gov/cdo-web/search}} (data is retrievable by filling in the forms), FTP\footnote{\url{https://www1.ncdc.noaa.gov/pub/data/ghcn/daily/}} (data is retrievable by dump downloading), and Service APIs\footnote{\url{https://www.ncdc.noaa.gov/cdo-web/webservices}} (data is retrievable by programmable methods). We conclude from employing the REST APIs for obtaining the NOAA climate data as they are made up of parameterized HTTP methods that are generally meant for developers. As compared with the former two services, REST API is the most frequently changing, lightweight, and least constrained, typically beneficial to be utilized as the building block for creating the climate observations KG in this study. 
 
 
\subsubsection{Upper-level semantics creation}
\label{section: ontTrans}
A notable characteristic of the KG is that it is structured under some domain specific ontology, \textit{i.e.} a collection of human- and machine-readable vocabularies for describing concepts and relationships inferred by data~\citep{gruber2018ontology}. Ontology confers data a semantic manifestation of their relationships~\citep{hogan2020knowledge}. 
The main advantage is that it allows for the creation of an accurate machine-understandable data structure, allowing data to be processed and inferred by algorithms and logic (knowledge reasoning) in a manner comparable to human thinking~\citep{707688}.  The CA ontology reuses parts of certain additional open ontologies, \textit{e.g.} SSN/SOSA ontology (sensors modeling), WGS84 ontology\footnote{\url{https://www.w3.org/2003/01/geo/}} (spatial entities modeling), QUDT ontology\footnote{\url{http://www.qudt.org/}} (units of measure, quantity kinds, and data types modeling), to organize NOAA climate data obtained over HTTP to develop the semantics of the climate KG. The reuse of existing ontologies can lower the cost of developing an ontology for a given topic and improve the interoperability between computer applications. 

\begin{figure}[ht]
\centering
\includegraphics[width=0.8\textwidth]{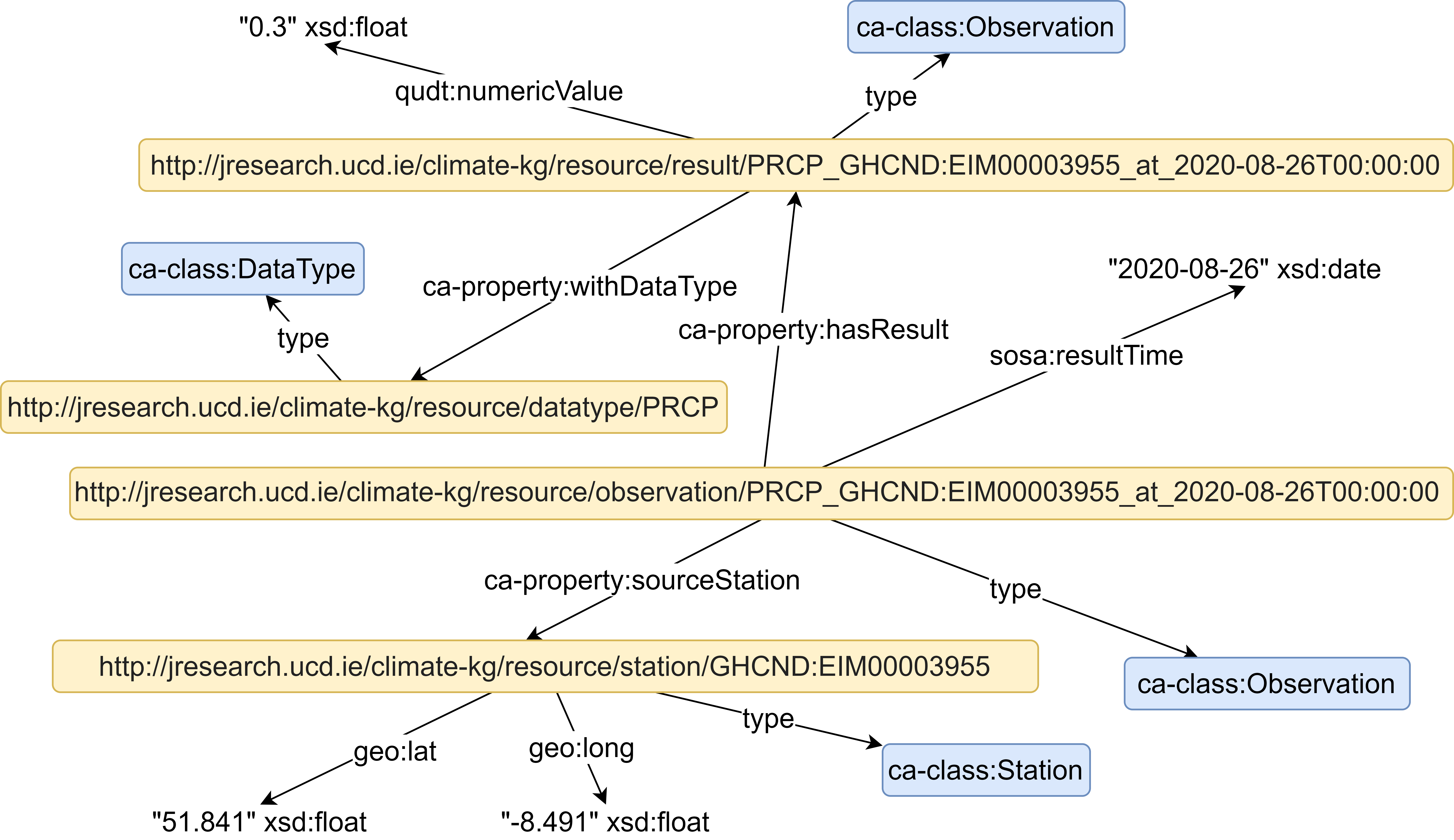}
\caption{A sub-graph about semantic representation of a precipitation observation resulted from a station in Cork Ireland}
\label{fig:observation}
\end{figure}

The vocabulary used in CA will correspond with the data schema underpinning the data acquired via the NOAA Service APIs, as well as providing more human-understandable interpretations for the data by a simple subject-predicate-object grammar (\textit{e.g.} ``Dublin''-``isCityOf''-``Ireland''). 
A summary of the steps involved in the entire procedure is as follows. (1) The first step is to look at the structure of the NOAA climate data format. We deduced the schema indirectly from a number of Service APIs' HTTP endpoints because NOAA does not provide it explicitly. This is theoretically feasible since NOAA climate data records of various categories are delivered as key-value pairs with a predetermined set of keys depending on the endpoint. For instance:  ``\texttt{\{baseurl\}/datasets}'' offering access to dataset instance records only (\textit{e.g.} ``Global Summary of the Year'' dataset), and ``\texttt{\{baseurl\}/stations}'' available only for station instance records. Furthermore, these endpoints accept utilizing arguments to impose scope constraints for data selection. For instance, request to ``\texttt{\{baseurl\}/stations?locationid=FIPS:UK}'' will simply return all UK's stations information. The above API features imply a clear data structure that is conducive to KG schema alignment. After that, (2) we define a collection of vocabularies that are inspired by the aforementioned accessible HTTP endpoints and their accompanying retrievable climate data entries to establish the ontology (CA) for the proposed climate KG. These vocabularies are grouped into two categories: \textit{Classes} and \textit{Properties} each of which will only include a few of the most relevant keywords in this paper. (Fig.~\ref{fig:observation} shows a example of how CA ontology describes the NOAA climate data).

\paragraph{Classes} The main classes defined in CA are consistent with the number of endpoints used in Service APIs. Table \ref{table:ontClass} lists in rows the alignment of our CA classes with the NOAA Web Service APIs endpoints, as well as their interpretations.

\begin{table*}[h!]
\caption{CA Ontology Classes vs. NOAA API endpoints}
\label{table:ontClass}
\begin{center}
\begin{tabular}{ | p{3cm} | p{3cm}| p{5cm} | } 
\hline
Ontology Term & Endpoint & Interpretation \\ 
\hline
Dataset & /datasets & Class for dataset instances (\textit{e.g.} ``GSOY'', ``GHCND'') \\ 
\hline
DataCategory & /datacategories &  Class for data category instances (\textit{e.g.} ``Precipitation'', ``Temperature'') \\
\hline
DataType & /datatypes & Class for data type instances (\textit{e.g.} ``Average Temperature'') \\
\hline
LocationCategory & /locationcategories & Class for the category of location (\textit{e.g.} ``City'', ``Country'') \\
\hline
Location & /locations & Class for locations (\textit{e.g.} ``Dubai, AE'') \\
\hline
Station & /stations & Class for stations \\
\hline
Observation (SOSA) & /data & Class for observations made by stations\\
\hline
\end{tabular}
\end{center}
\end{table*}

\paragraph{Properties} The main ontology properties are aligned with the fields of retrieved JSON data from the NOAA API endpoints and can be referred to on Table \ref{table:ontPrpt}.

\begin{table*}[h!]
\caption{CA Ontology Properties vs. Keys of Retrieved Json data}
\label{table:ontPrpt}
\begin{center}
\begin{tabular}{ | p{3cm} | p{3cm}| p{5cm} | } 
\hline
Onotlogy Term & Data Field & Interpretation \\ 
\hline
name & ``name'' &  literal name of an entity \\
\hline
isLocatedIn & parameter ?locationid on request & location of an entity (if applicable)  \\
\hline
elev/lat/long & ``elevation''/``latitude''
/``longitude'' & elevation/latitude/longitude of an entity (if applicable) \\
\hline
elevUnit & ``elevationUnit'' & the measure units of entity's elevation \\
\hline
inDataCategory & parameter ?datacategoryid on request & the data category of an data type \\
\hline
sourceStation & ``station'' & the station where the data release \\
\hline
resultTime (SOSA) & ``date'' & the time when observation happened \\
\hline
withDataType & ``datatype'' & the data type of the data value \\
\hline
numericValue (qudt) & ``value'' & the observed value \\
\hline
hasResult (SOSA) & nested with withDataType, numericValue & the observed result \\
\hline
\end{tabular}
\end{center}
\end{table*}

\subsubsection{Ontology alignment} \label{sec:ontalign}
The CA ontology is built using a blend of the vocabulary from open ontologies and the vocabulary we created independently. In order to achieve better integrative compatibility (providing the identical semantics) with large outside linked datasets, we align some of the classes and properties we created exclusively in CA with vocabularies that are either already accepted as extendable ontologies by W3C (World Wide Web Consortium\footnote{\url{https://www.w3.org/}}) standards, such as \textit{SOSA/SSN}, or defined in sophisticated KGs owned by other communities. In this study, ontology alignment was accomplished manually by identifying the conceptual relations between CA and other ontologies, such as concept inclusion and concept equivalence. A key benefit of manual alignment in this research is guaranteeing the correctness of the ontology alignment from an expert's perspective for minimizing non-standard ontologies and maximizing the usage of standard ontologies such as \textit{SOSA/SSN}. For any more studies that want to adapt and scale up this methodology, it may be helpful to adopt some automated ontology alignment methods initially, followed by expert review, in order to reduce the alignment time for ontological heterogeneity while ensuring the correctness.

The alignment generally happens when associating CA ontology concepts (\textit{i.e.} class and property terminologies) to those established in other ontologies using RDF schema\footnote{\url{https://www.w3.org/TR/rdf-schema/}} and OWL (the Web Ontology Language) vocabularies\footnote{\url{https://www.w3.org/TR/2012/REC-owl2-overview-20121211/}}. 
For example, in CA, \textit{rdfs:subClassOf} is applied to \textit{ca:Result} to extend the W3C standard class \textit{sosa:Result}, allowing the related instances to be assigned the CA-specific properties. Furthermore, because several KGs in this field have specified a number of well-conceptualized vocabularies, we match some of the CA vocabularies with the ontologies applied to those KGs by finding vocabularies with equal meanings. This study targets the Spanish climate KG—AEMET\footnote{\url{http://aemet.linkeddata.es/index_en.html}}, and alignment between all identical conceptions is achieved using \textit{owl:sameAs} (for example, \textit{ca:Location$\rightarrow$owl:sameAs $\rightarrow$aemet:AdministrativeArea}, \textit{ca:Station$\rightarrow$owl:sameAs $\rightarrow$aemet:WeatherStation}). This alignment will allow for the formation of conceptional linkages to the AEMET KG, as well as the avoidance of lexical repetition caused by the development of similar concepts in different name-spaces.

\subsubsection{Collection strategy} \label{sec:clctstra}
On the NOAA website, climate data is updated using uploads from real-world observation stations. However, because data is transmitted at different times from different stations, collecting all of the updates at the same timestamp is impractical.  
To compensate, this study employs a simple but effective data gathering approach that downloads data from a fixed-length historical period (also known as a sliding window) prior to the data gathering time. Every historical record will undoubtedly be duplicated several times as the collection progresses, until the historical records run out of the window. Although employing a long duration length improves the completeness of fetched data, it is evident that this benefit is contingent on the amount of repeated data growing with the window length. We empirically settle on a four-week timeframe, which is long enough to minimize data loss owing to submission delays, while also avoiding excessive data collection duplication. It is also worth noting that duplication issue can be skipped loosely when inserting data into a KG, because any triplestore that complies with the \textit{Graph Store HTTP Protocol}\footnote{\url{https://www.w3.org/TR/sparql11-http-rdf-update/}} would delete duplicates effectively.

\subsubsection{Publishing as Linked Data}\label{sec:publ}
Data published as Linked Data is key to efficiently constructing KGs. The core notion of Linked Data is to create a network of data in which every single piece of published data on the Internet has the ability to connect to and be connected to other Linked Data sources~\citep{orlandi2019interlinking}. This work publishes CDO datasets as Linked Data, allowing users to take advantages of HTTP technologies to investigate the possibilities of intersecting data with domains other than climate. One of the key benefits of this approach is the capability of enriching the data context, as seen in Section~\ref{sec: linkage to wiki}. 

We setup a RDF triplestore for the storing of climatic RDF triples and also configure the Web-wide SPARQL endpoint\footnote{\url{http://jresearch.ucd.ie/kg/}} to establish NOAA  climate data as Linked Data. Researchers that are interested in these data may use the endpoint to submit queries directly. Obtaining data from a SPARQL endpoint, on the other hand, necessitates some understanding of the SPARQL language, which may be unfamiliar to researchers from various backgrounds who have no prior experience with Linked Data processing. To unravel the complexity in the use of SPARQL to some extent, a step-by-step online guide --which will be elucidated in Section~\ref{sec:web interface}-- is created for the first user interactions with the SPARQL query language on exploring NOAA climate data. 

\subsubsection{Linkage to OpenStreetMap and Wikidata}
\label{sec: linkage to wiki}
In many situations of climate-related research, such as these more recent studies~\citep{stockhause2020documentation, vermeulen2020supporting}, 
data from several domains is likely to be collected to see how climate interacts with other domains (\textit{e.g.} global warming vs. urbanization). This generally entails analysis covering a wider range of data sources. As Linked Data is one of popular ways for this sort of analysis~\citep{zouaq2017linked}, we explore more in this area to connect more supplementary data with the proposed NOAA linked climate data. The general tactics proposed in this paper is to achieve the linkage of the NOAA climate data with OpenStreetMap nodes and Wikidata which has recently been discovered to be the most competitive open encyclopedic data source of detailed information about countries, cities, persons, among others (\textit{e.g.} DBpedia\footnote{\url{https://wiki.dbpedia.org/}}, YAGO\footnote{\url{https://yago-knowledge.org/}})~\citep{mora2019systematic}.

\begin{figure}[ht]
\centering
\includegraphics[width=0.8\textwidth]{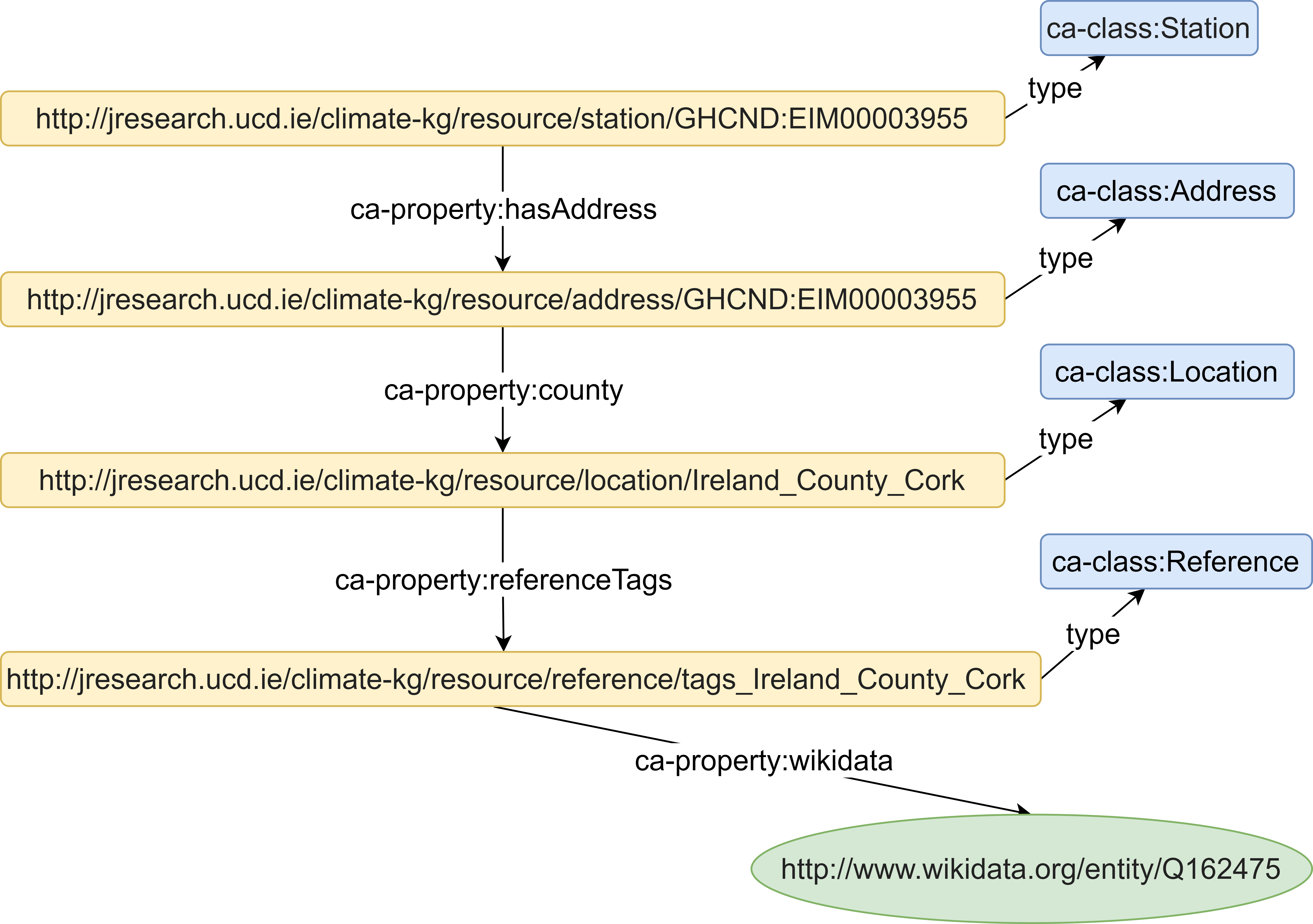}
\caption{A sub-graph about semantic representation of a station in Cork Ireland}
\label{fig:station}
\end{figure}

To implement the aforementioned linkage, we begin with all stations recorded in CDO datasets in terms of their explicitly recorded longitude and latitude coordinates. Then, over HTTP, OpenStreetMap is employed to request detailed geographical information--country subdivisions that include the coordinates--via its reverse geocoding APIs\footnote{\url{https://nominatim.org/release-docs/develop/api/Reverse/}}. The returned results include rich subdivisions in relation to the administrative areas of the coordinates from levels of country, city, county etc. of which each entity is bound with its unique WikiData code. Finally, these geographical data, which are stored in various formats (e.g. JSON), are transformed into CA ontology-compliant data using a well-designed mapping process, and then linked to the station that generated them in the first place (a graphical view of this linkage is given in Fig.~\ref{fig:station}). As a result, our Linked Data platform now supports joint queries combining Wikidata for contextual information regarding a geographic node. For a query example, see Listing 1, where ``?sta'', ``?wb'', ``?addr'', ``?loc'', ``?wd'' represent station, water body, address, location, and Wikidata key, respectively. Property wdt:P206 is defined by Wikidata ontology as the interpretation ``located in or next to body of water''?. The solutions to this query in effect answer the nearby water bodies (\textit{i.e} oceans, seas, lakes etc.) of all stations of concern in CDO datasets, which can be helpful to find the weather discrepancies caused by water bodies' characteristics.

\begin{figure*}[htb]
\centering
\includegraphics[width=0.97\textwidth]{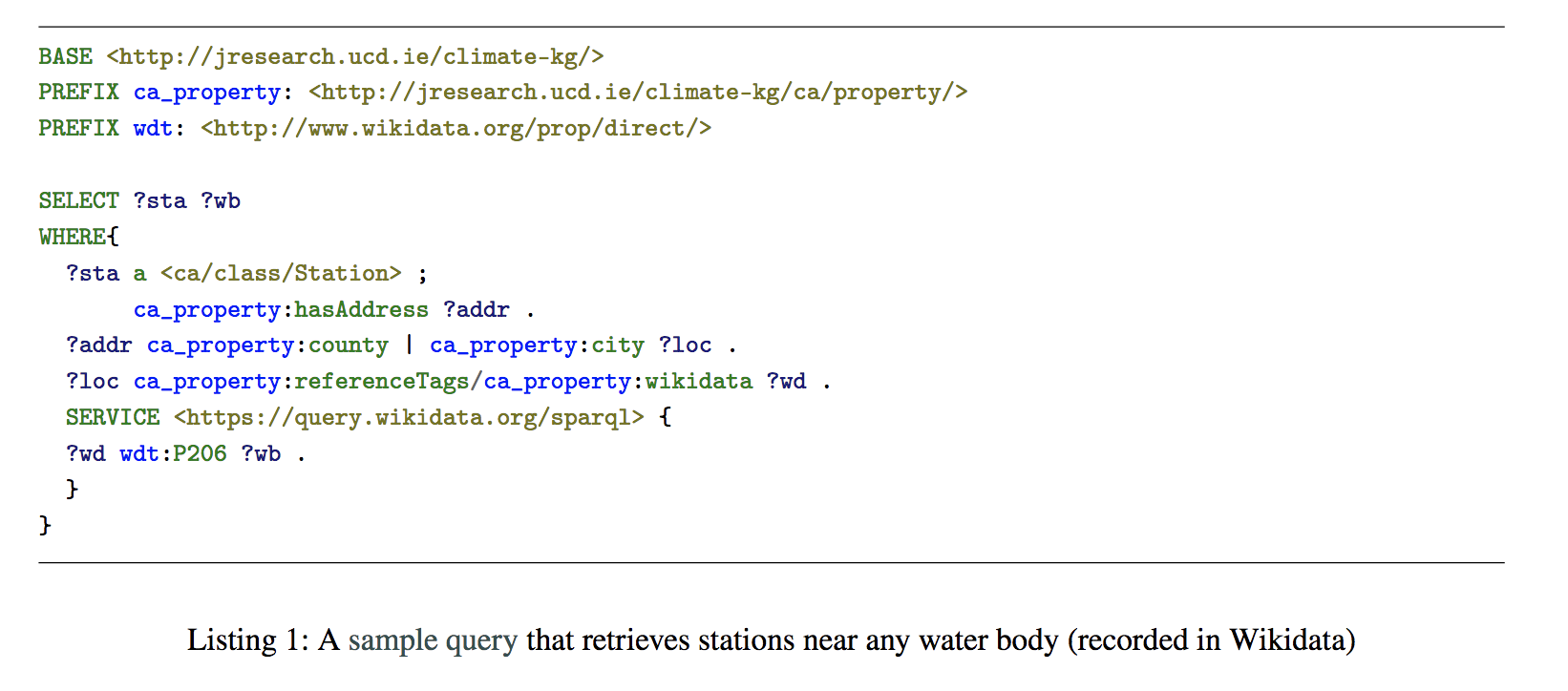}
\end{figure*}



\subsubsection{URI dereference}
\label{sec:uri deref}

\begin{figure}[ht]
\centering
\includegraphics[clip, trim=0.0cm 5cm 0cm 5cm, width=0.8\textwidth]{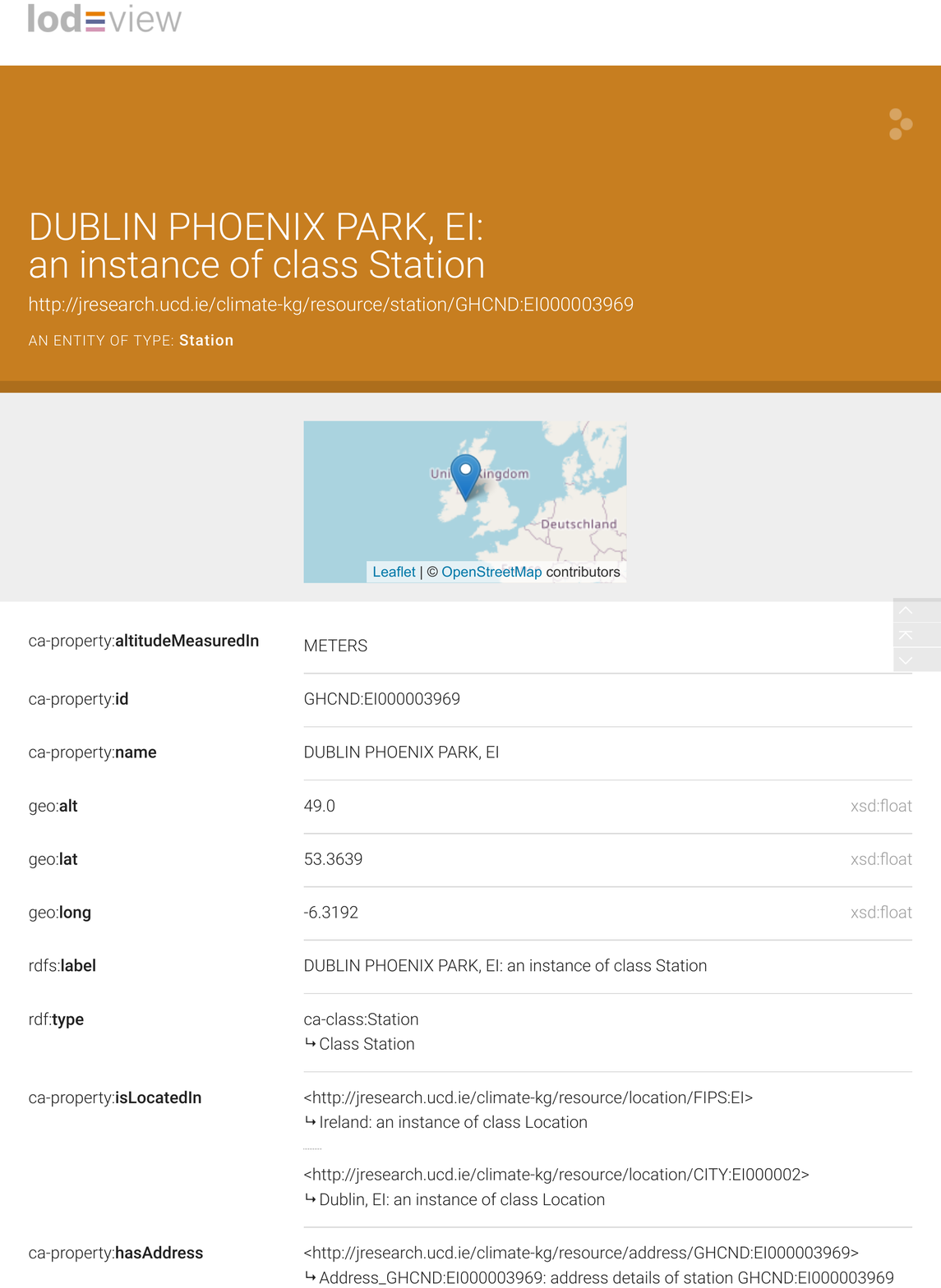}
\caption{Dereferencing to \url{http://jresearch.ucd.ie/climate-kg/resource/station/GHCND:EI000003969} }
\label{fig:dereference}
\end{figure}

This work supports URI deferencing by providing precise RDF statements for each URI-named entities in the SPARQL endpoint. Users might read the statements to comprehend the semantic meaning associated with the URI-denoted item. When related entities (\textit{e.g.} ``Ireland'') and literal properties (\textit{e.g.} ``altitude'', ``longitude'', and ``latitude'') are presented, the URI identifying the ```DUBLIN PHOENIX PARK'' station is specified by additional associated information, as illustrated in Fig~\ref{fig:dereference}. Furthermore, users of this dereferencing system have access to an additional graphical view of the knowledge network in which graph nodes (\textit{i.e.} URIs) are interconnected with other nodes and the scope of the view may be expanded along with node access.

\subsection{Implementation}

In practice, the above workflow is built by a set of Python scripts which are available in our Github repository\footnote{\url{https://github.com/futaoo/LinkClimate}}. In summary, these scripts implements functionalities including (in order): definition of the proposed CA ontology, weekly retrieval of online NOAA climate data, data transformation, and lastly storage as Linked Data. We now have 14 millions RDF triples about climate data stored in our Jena/Fuseki SPARQL endpoint.

\subsection{Knowledge graph evaluation with competency questions}\label{sec:eval}
Competency questions~\citep{Debruyne2022-fi} are a collection of questions proposed by a group of domain stakeholders for the purpose of evaluating a system's capacity to answer these questions using the data it maintains. If the system is capable of answering the competence questions, the information is considered acceptable and well-organized (\textit{by the ontology}) for the domain applications. The LinkClimate knowledge graph is designed to accommodate the climate data requirements for potential analysis. To evaluate LinkClimate in terms of possible climate analytical scenarios in the domain,  we convened multiple meetings with domain experts and compiled the following 11 competency questions for LinkClimate to address. 

\begin{enumerate}[label=\arabic*.]
    \item \textit{Where are all the stations that are located in a particular administrative region?}
    \item \textit{Which station is the nearest to a certain station?}
    \item \textit{Which stations fall inside a certain range of latitude and longitude coordinates?}
    \item \textit{How can I group stations according to a particular observed climatic variable?}
    \item \textit{How to find climate variables that are recorded by a particular station?}
    \item \textit{How long has a certain climatic variable been monitored by a particular station?}
    \item \textit{How can a time series for a single climatic variable be retrieved?}
    \item \textit{How can a time series for a number of different climatic variables be retrieved?}
    \item \textit{How can station observations be aggregated according to their temporal resolution?}
    \item \textit{How can I determine the station's geographical context?}
    \item \textit{How to include extra environmental data into a station's or set of stations' knowledge graph in order to do cross-domain data analysis?}
\end{enumerate}

The knowledge graph is evaluated using the approach proposed by Haussmann et al. in the paper~\citep{Haussmann2019-yo}, which involves querying the knowledge graph's underlying ontology using SPARQL. Because the knowledge graph's information, such as the links between stations and climatic variables, is directly reflected in the ontology (see Section~\ref{section: ontTrans}), the evaluation also reveals the ontology's competence for the domain applications. The evaluation results indicate that SPARQL queries are capable of providing adequate responses to competency questions. Specifically, for questions concerning the geographical relationships between stations, \textit{i.e.}, questions 2 and 3, SPARQL can currently provide answers by querying the longitude and latitude coordinates. To determine the geographical relationships (\textit{e.g.} locating stations within a given area such as a bounding box in terms of south Latitude, north Latitude, west Longitude, east Longitude), however, additional computations based on the coordinates must be conducted. To further simplify this type of geographical computing using SPARQL, we want to include the GeoSPARQL semantics with an engine implementation~\citep{Car2022-rs} into our framework in the future, allowing for more queryable geographical relations. The remaining questions can be answered directly.  To summarize, we present examples of SPARQL queries written for questions 8 and 9. We advise visitors to this page\footnote{\url{https://github.com/futaoo/LinkClimate/blob/main/Competency-Questions.md}} for an in-depth evaluation process. In summary, each of the competence questions above can be addressed using a SPARQL query, and LinkClimate supplies the expected data for all of them.  

\begin{figure*}[htb]
\centering
\includegraphics[width=0.97\textwidth]{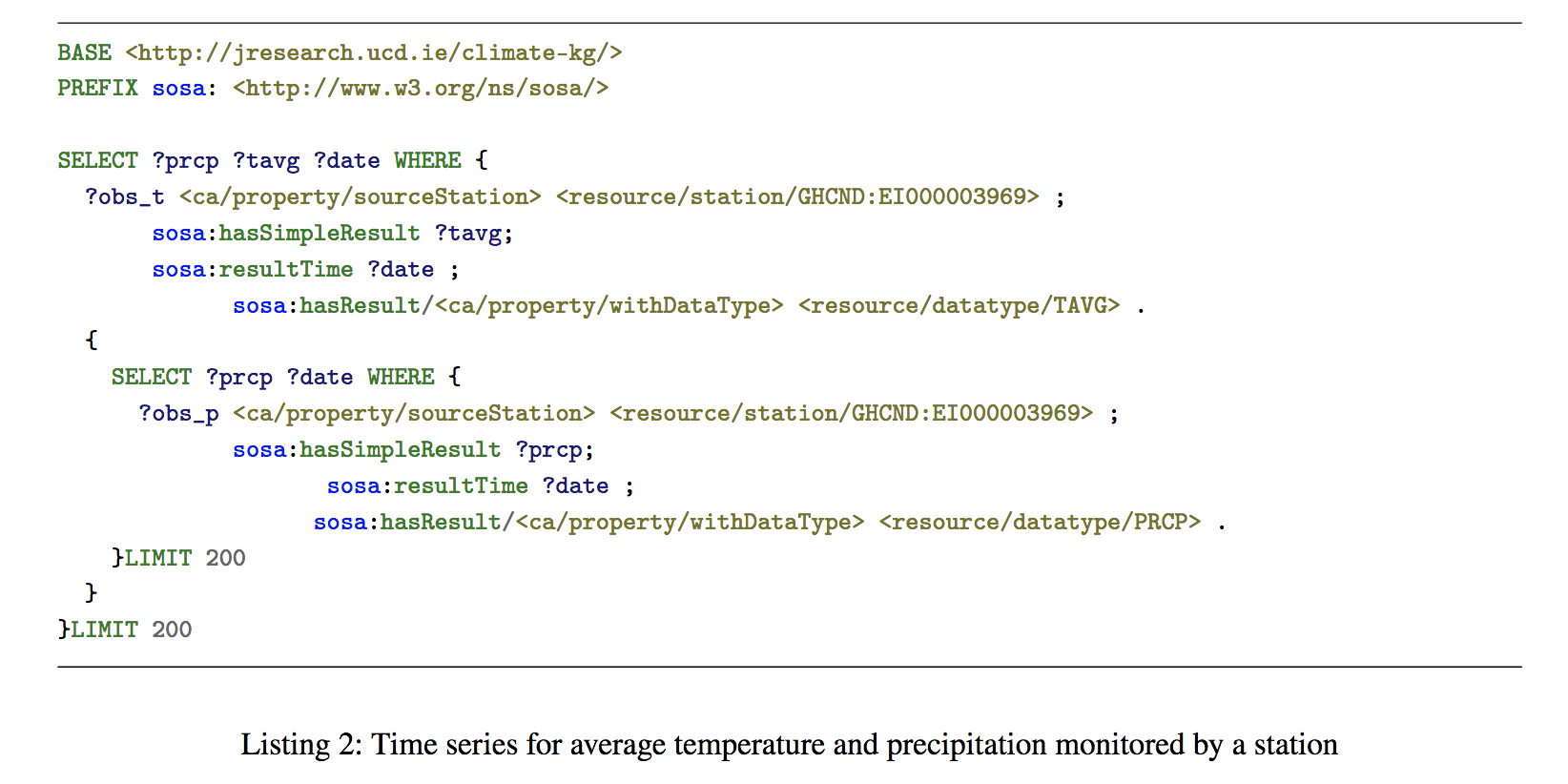}
\end{figure*}

\begin{figure*}[htb]
\centering
\includegraphics[width=0.97\textwidth]{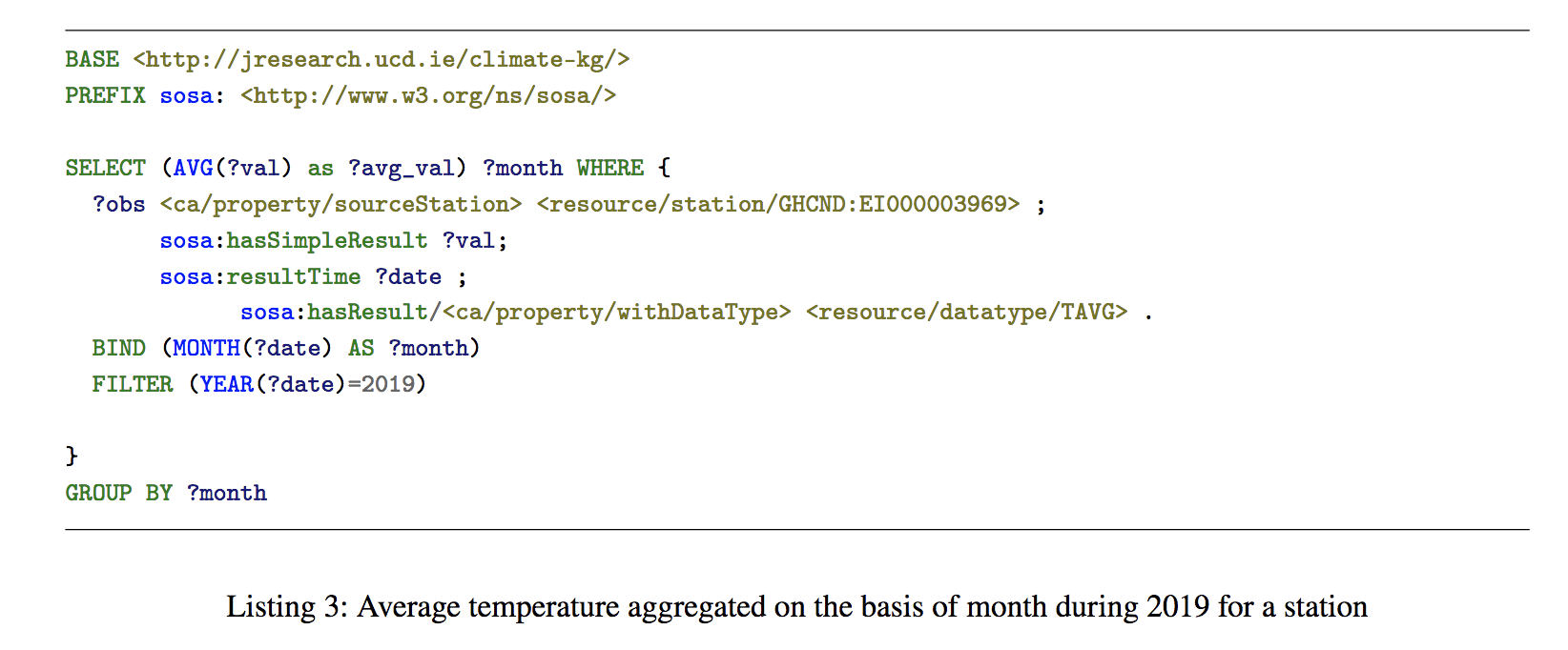}
\end{figure*}

\section{Web Interface}
\label{sec:web interface}
We created a supplementary instructional website\footnote{\url{http://jresearch.ucd.ie/linkclimate/}} for users who are unfamiliar with the KG database and its involved operations (\textit{e.g.} SPARQL query). Users can learn about the CA ontology through schematic diagrams presented on the website and then have some quick hands-on queries to our SPARQL endpoint by following a step-by-step beginner's guidance. Third-party tools used in this project such as Jena Fuseki, are also detailed on the website so that users can learn how to utilize them. Finally, a usability test has been undertaken to obtain first-hand feedback from testing participants about their experience on using the website. 

\subsection{Interface Overview}
\label{sec:interfaceoverview}
\begin{figure}[ht]
\centering
\includegraphics[width=\textwidth]{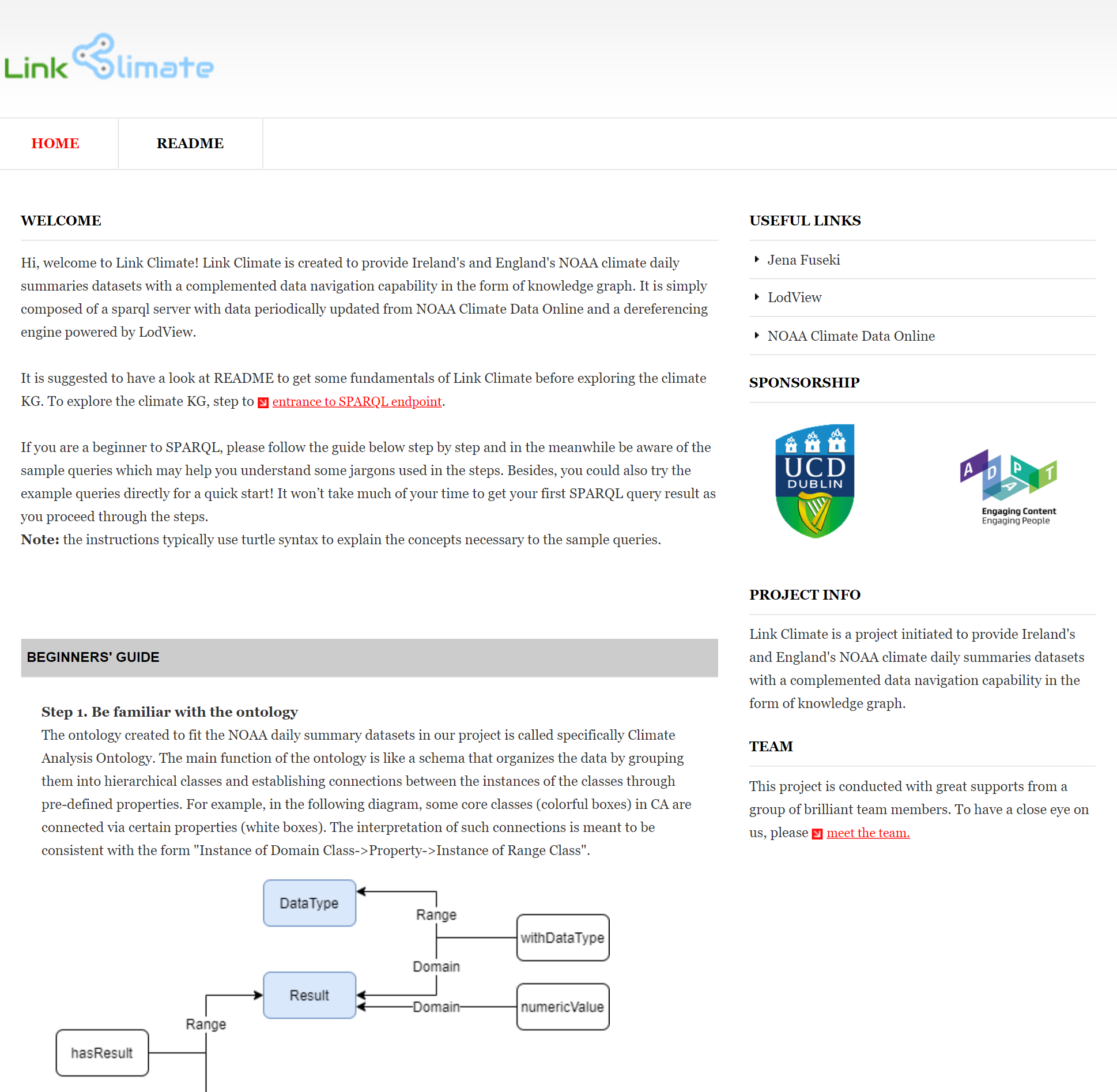}
\caption{A screenshot of our web interface.}
\label{fig:webui}
\end{figure}

The website's goal is to assist users in discovering and operating our climate  KG (technique perspectives) and propagate KG applications in the climate domain (project perspectives). The website (a snapshot is shown in Fig.~\ref{fig:webui}) primarily gives instructions based on two subjects, one of which is a ``Readme'' instruction and the other is a ``Beginners' guide''. The purpose of the ``Readme'' is to familiarize users with the CA ontology's key vocabulary and how the structure of climate data is modeled. Beginners should use the KG as a starting point for their practice by following the tasks specified by ``Beginners' guide''. Users are encouraged to read the ``Readme'' and ``Beginners' guide'' to gain a basic understanding of the climate  KG and to get some hands-on practice. The usability test in Section~\ref{sec: usability test} will show how this website design meets the desired goals.

\subsection{Usability Test}
\label{sec: usability test}
The purpose of usability testing is to confirm that the product can work according to the specification. A successful usable product is supposed to guarantee that the product can help users to complete their tasks in a manner which easily meet their needs~\citep{dumas1999practical}. We designed tasks for potential users to practice and gather their feedback after traversing the content of the web interface. This allows us to have a more in-depth understanding of the success of each functionality, as well as thoughts on how to improve the web interface in the future iterative development on the interface. 

\subsubsection{Methods}\label{sec:methd}
We perform a usability test to obtain feedback on how the features of the web interface satisfy users in exploring our climate  KG. Our Human Research Ethics Committee has authorized the test with code (LS-E-20-199-Wu-Dev). The participants were recruited using advertisements on several social media platforms (\textit{e.g., Twitter, WeChat, and a standalone Web-page}) as well as emails that included a link to a Google form for collecting replies. The majority of the participants have little experience with KGs but have a basic comprehension of computer science fundamentals, such as database fundamentals. For people who are not proficient in English, a translated survey form (Chinese version) is supplied for helping them readily give the answers. In total, we have received 31 user responses about their practice with the web interface.

The protocol of the questionnaire is designed referring to the widely used PSSUQ (Post-Study System Usability Questionnaire)~\citep{lewis1992psychometric} which particularly takes into account together the usage of pre-designed tasks and the usability of the website~\citep{Fruhling2005-qq,Valadi_undated-xy} amongst other questionnaire standards, such as the SUS (Software Usability Scale), QUIS (Questionnaire for User Interaction Satisfaction), and SUMI (Software Usability Measurement Inventory). During the usability test, participants are asked to finish a number of SPARQL query tasks posted on the website before completing the questionnaire. The questionnaire contains the following 7 Likert scale questions (1 lowest → 5 highest) and a blank area in the end for participants to leave their comments. 


\begin{enumerate}[label=Q\arabic*.]
    \item \textit{I would agree that the website could properly spark users' interest in getting their hands on various functions integrated on the website.}
    \item \textit{I am clear about the topic of the project after reading all the descriptive information on the webpage.} 
    \item \textit{I thought the beginner's guide could reduce the difficulty for non-expert users in accessing some climate data from a KG.}
    \item \textit{I have a preliminary knowledge of how to get some climate data after exploring the website.}
    \item \textit{I have a preliminary knowledge of how to navigate the climate data by avail of the provided GUI (step 4 in beginner's guide).}
    \item \textit{With the GUI, I would have a more comprehensive understanding of the knowledge graph structure.}
    \item \textit{I am confident that the project is favorable to people who work with climate data. }
\end{enumerate}
The questions are designed to fall into 2-class categories, as mentioned in Section~\ref{sec:interfaceoverview}: project level and technical level. At the project level, users are asked to rate \textit{Curiousness} (Q1), \textit{Understandability} (Q2), and \textit{Usefulness} (Q7) in terms of their experience on the application of KG in the climate domain. At the technical level, we aim at \textit{Instructiveness of SPARQL} (Q3 and Q4), and \textit{Instructiveness of GUI} (Q5 and Q6), which are two important essentials given by this website. The following Section~\ref{sec:result} examines the overall evaluation results and users' comments.
\subsubsection{Results}
\label{sec:result}
\begin{figure}[t!]
\centering
\includegraphics[width=\textwidth]{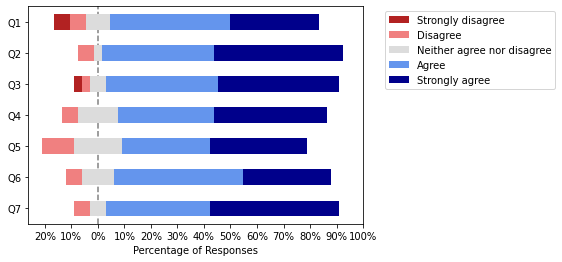}
\caption{Percentage of responses according to questions}
\label{fig:respons}
\end{figure}

We first calculated the mean score (M) and standard deviation (STD) of each question, as shown in Table~\ref{table:stat}, then presented the overall Likert responses in a single diagram (see Fig~\ref{fig:respons}) to provide a visual perception of the statistics made on the scores graded by users. In general, the combination of Table~\ref{table:stat} and Fig~\ref{fig:respons} demonstrates a positive attitude is received from participants towards our Web interface, both at project level and technical level. In particular, Q2 and Q7 reaches the highest average scores (M$>$4.30, STD$<$0.85) hence the web interface is most effective in making users easily understand the initiative of our climate KG project and is considered as a useful tool to boost the climate research in relation to climate data. For the other aspects in testing, \textit{i.e.} \textit{Curiousness}, \textit{Instructiveness of SPARQL}, and \textit{Instructiveness of GUI}, the answers are generally around (M$\approx$4.00, STD$\approx$1.00) which guarantee that the instructive tasks are acceptable in practice. However, some users who gave lower scores also left the following comments which are valuable for directing our future developments:

\begin{enumerate}[label=(\arabic*)]
    \item \textit{``I think it is an attractive project, with very concise tutorials and case demonstrations...but I may be a little strange to GUI...I tried to read the text for several times...''}
    \item \textit{``...the `Beginner's Guide' and `Sample Queries' sections talk more about SPARQL instead...a good idea to smoothen the transition by letting him use the tool first (without any technicalities) and then gradually unfold the technical aspects.''} 
    \item \textit{``In order to better spark users' interest...it would be great to have an embedded interface on the home page, so that users could play around.''}
    \item \textit{``It would likely be more helpful if the Beginner's Guide section showed both query and sample output rows as illustrations, to provide more context for what's being described.''}
    \item \textit{``...Clicking on the "Try it on Sparql" should probably open a new tab/window as it is an example...''}
\end{enumerate}

The comments reflect that the GUI and SPARQL instructiveness should be more intuitive, and one of the most shared user expectations is to embed part of the application in a frame directly on the website, instead of using redirecting links. Additionally, the order of instruction and visualization is advised to be altered so to raise users' interest immediately with a visualization tool and then unfolding the technical specifications later.

\begin{table}[t!]
\setlength\tabcolsep{0pt} 
\caption{A summary of mean and standard deviation calculated on the score obtained of each question} \label{table:stat} 
\sisetup{round-mode=places,round-precision=2} 
\begin{tabular*}{\textwidth}{@{\extracolsep{\fill}} l *{14}{S[table-format=-1.2]} }
\toprule
 & {Q1} & {Q2} & {Q3} & {Q4} & {Q5} & {Q6} &{Q7} \\
\midrule
M & 3.94 & 4.33 & 4.24 & 4.15 & 3.94 & 4.09 & 4.30  \\
STD & 1.12 & 0.82 & 0.94 & 0.91 & 1.03 & 0.84 & 0.85 \\
\bottomrule
\end{tabular*}
\end{table}

\vspace{1.5cm}
\section{Conclusions and Future Work}
This paper proposes an approach that targets online climate data, specifically NOAA data, which contains extensive archives of 1) historical climatic observations of meteorological variables and 2) meteorological stations from throughout the world. We fetch these data archives and construct a climate  KG to store them. The KG is then automatically synchronized with the dynamic data sources and published on an Open Linked Data platform, publicly available on the Web. This facilitates the local KG's extensibility, as new datasets that observe the Linked Data protocol may be retrievable via linking with identical data instances in the Linked Data platform. As all of the stations in NOAA datasets are specified by latitude and longitude, we gather more information regarding the geographic coordinates by searching in the OpenStreetMap database. Especially, the Wikidata codes of the coordinates are used to introduce the Wikidata descriptions according to the coordinates in our climate  KG. As a result, users can explore more information on the stations, beyond the NOAA online climate data, for example, identifying nearby water bodies, mountains, etc. All this information is presented to users via a Web interface exposing Linked Data in a dynamic and transparent way. Finally, we conduct a survey with real users to collect their feedback on their experience with our platform. The results show that the platform performs well in unraveling the difficulties non-expert users may encounter when they make attempts to explore the KG, and the instructions provide fundamentals of climate  KG. In the future, we will first start taking into account the advice from users' feedback, embedding applications directly in a frame on the website to strengthen the instructiveness of the climate KG platform. Then, the next phase will be to add more concepts into CA ontology for the integration of other climate-related data sources, such as remote climate data sources, such as air pollution, ocean data and NetCDF formatted data. Depending on the complexity level of the extended CA ontology, we may further apply a semi-automatic ontology alignment approach to see if it can actually expedite the establishment of compatibility with other ontologies. The anticipated outcome will boost the cross-domain climate related research\citep{Zeng2019-ek}. 

\section*{Acknowledgments}
This research was partially funded by the EU H2020 research and innovation programme under the Marie Skłodowska-Curie Grant Agreement No.~713567 at the ADAPT SFI Research Centre at Trinity College Dublin. The ADAPT Centre for Digital Content Technology is funded under the SFI Research Centres Programme (Grant 13/RC/2106\_P2) and is co-funded under the European Regional Development Fund.

\section*{Computer Code Availability}
The source code related to this manuscript is available via \texttt{\url{https://github.com/futaoo/LinkClimate}}.

\section*{CRediT author statement}
Jiantao Wu: Conceptualization, Methodology, Software, Data curation, Writing- Original draft preparation, Visualization, Investigation, Writing- Reviewing and Editing; Fabrizio Orlandi: Conceptualization, Methodology, Software, Data curation, Supervision, Writing- Original draft preparation, Investigation, Validation, Writing- Reviewing and Editing; Declan O'Sullivan: Methodology, Writing- Original draft preparation, Writing- Reviewing and Editing; Soumyabrata Dev: Conceptualization, Methodology, Supervision, Writing- Original draft preparation, Investigation, Validation, Writing- Reviewing and Editing.

\balance

\end{document}